\documentclass[aps,rmp,twocolumn,showpacs,a4paper,superscriptaddress]{revtex4}
\usepackage{graphicx}
\usepackage{amsmath,amsfonts,amssymb}
\renewcommand{\l}{\lambda}

\renewcommand{\d}{\partial}

\newcommand{\dd}{{\rm d}}

\newcommand{\be}{\begin{equation}}
\newcommand{\ee}{\end{equation}}
\newcommand{\ba}{\begin{eqnarray}}
\newcommand{\ea}{\end{eqnarray}}

\begin{document}

\title{
Daubechies wavelets as a basis set for density functional pseudopotential calculations}
\author{Luigi Genovese}
\affiliation{Institut de Nanosciences et Cryog\'enie,\\
         SP2M/L\_Sim, CEA-Grenoble, 38054 Grenoble cedex~9, France}
\author{Alexey Neelov}
\affiliation{ Institut f\"{u}r Physik, Universit\"{a}t Basel, Klingelbergstr. 82, 4056 Basel, Switzerland}
\author{Stefan Goedecker}
\affiliation{ Institut f\"{u}r Physik, Universit\"{a}t Basel, Klingelbergstr. 82, 4056 Basel, Switzerland}
\author{Thierry Deutsch}
\affiliation{Institut de Nanosciences et Cryog\'enie,\\
         SP2M/L\_Sim, CEA-Grenoble, 38054 Grenoble cedex~9, France}
\author{Seyed Alireza Ghasemi}
\affiliation{ Institut f\"{u}r Physik, Universit\"{a}t Basel, Klingelbergstr. 82, 4056 Basel, Switzerland}
\author{Alexander Willand}
\affiliation{ Institut f\"{u}r Physik, Universit\"{a}t Basel, Klingelbergstr. 82, 4056 Basel, Switzerland}
\author{Damien Caliste}
   \affiliation{Institut de Nanosciences et Cryog\'enie,\\
         SP2M/L\_Sim, CEA-Grenoble, 38054 Grenoble cedex~9, France}
\author{Oded Zilberberg}
\affiliation{ Institut f\"{u}r Physik, Universit\"{a}t Basel, Klingelbergstr. 82, 4056 Basel, Switzerland}
\author{Mark Rayson}
\affiliation{ Institut f\"{u}r Physik, Universit\"{a}t Basel, Klingelbergstr. 82, 4056 Basel, Switzerland}
\author{Anders Bergman}
\affiliation{Institut de Nanosciences et Cryog\'enie,\\
         SP2M/L\_Sim, CEA-Grenoble, 38054 Grenoble cedex~9, France}
\author{ Reinhold Schneider}
\affiliation{Technische Universit\"at Berlin, Sekretariat MA 8-1, Str. des 17. Juni 136
D-10623 Berlin, Germany}

\begin{abstract}
Daubechies wavelets are a powerful systematic basis set for electronic structure calculations because they are orthogonal and localized both in real and Fourier space. 
We describe in detail how this basis set can be used to obtain a highly efficient 
and accurate method for density functional electronic structure calculations. 
An implementation of this method is available in the ABINIT free software package. This code shows high systematic convergence properties, very good performances and an excellent efficiency for parallel calculations.
\end{abstract}


\maketitle

\section{Introduction}
In recent years, the Kohn-Sham formalism of the density functional theory (DFT) approach has proven to be one of the most efficient and reliable first-principle methods for investigating material properties and processes that exhibit quantum mechanical behavior.
The high accuracy of the results together with the relatively simple form of the exchange-correlation functionals make this method arguably the most powerful tool for \textit{ab initio} simulations of the properties of matter.
The computational machinery of DFT calculations has been widely developed in the last decade, giving rise to a plethora of DFT codes. The use of DFT calculation has thus become more and more common, and its domain of application includes solid state physics, chemistry, material science, biology and geology. 

One of the most important characteristics of a DFT code is the set of basis functions used to express the Kohn-Sham (KS) orbitals. The domain of applicability of a code is tightly connected to this choice. For example, a non-localized basis set like plane waves is highly suitable for electronic structure calculations of periodic and homogeneous systems, while it is much less efficient in expanding localized functions, which have a wider range of components in reciprocal space. 
For these reasons DFT codes based on plane waves are not well suited to simulate inhomogeneous or isolated systems like molecules, due to the high memory requirements for such kind of simulations.

A strong distiction should also be made between codes that use systematic and non-systematic basis sets. A systematic basis set allows us to calculate the solution of the KS equations with arbitrarily high precision as the number of basis functions is increased. In other terms, the numerical precision of the results is related to the number of basis functions used to expand the KS orbitals. With such a basis set it is thus possible to obtain results that are free of errors related to the choice of the basis, eliminating a source of uncertainty. This is particularly important in view of the fact that highly accurate approximations to the exchange correlation functional are now available such as the PBE functional~\cite{pbe}. Some of these functionals also contain van der Waals interactions~\cite{vdw}.  A systematic basis set allows us to accurately calculate the solution of a particular exchange correlation functional. On the other hand, 
non-systematic basis sets, for example gaussians, often become over complete and numerical instabilities arise before absolute convergence can be achieved. Such basis sets are more difficult to use, since the basis set must be carefully tuned by hand by the user, which will sometimes require some preliminary knowledge of the system under investigation. This is the most important weakness of this popular basis set. 

Another property which has a role in the performances of a DFT code is the orthogonality of the basis set. The use of nonorthogonal basis sets requires the calculation of the overlap matrix of the basis functions and to perform various operations with this overlap matrix 
such as inverting the matrix, by iterative or non-iterative methods. This makes methods based on non-orthogonal systematic basis functions not only more complicated but also slower. 

Daubechies wavelets~\cite{daub} have virtually all the properties that one might desire of 
a basis set being used for the simulation of isolated or inhomogeneous systems. 
They form a systematic orthogonal and smooth basis, localized both in 
real and Fourier space and that allows for adaptivity.
A DFT approach based on such functions will meet both the requirements of precision and localization found in many applications. In this paper, we will describe in detail a DFT method based on a Daubechies wavelets basis set.
This method is implemented in a DFT code, distributed under GNU-GPL license and integrated in the \texttt{ABINIT}~\cite{abinit} software package.
A separate, standalone version of this code is also available and distributed under GNU-GPL license~\cite{bigdft}.
In the next few paragraphs we will discuss the importance of the properties of Daubechies wavelets in the context of electronic structure calculations.

A wavelet basis consists of a family of functions generated from a mother function and its translations on the points of a uniform grid of spacing $h$. The number of basis functions is increased by decreasing the value of $h$. Thanks to the systematicity of the basis, this will make the numerical description more precise.
The degree of smoothness determines the speed with which one converges to the exact result as $h$ is decreased. The degree of smoothness increases as one goes to higher order Daubechies wavelets. In our method we use Daubechies wavelets of order 16. This together with the fact that our method is quasi variational gives a convergence rate of $h^{14}$. Obtaining such a high convergence rate is essential in the context of electronic structure calculations where one needs highly accurate results for basis sets of acceptable size. 
The combination of adaptivity and a high order convergence rate is typically not achieved 
in other electronic structure programs using systematic real space methods~\cite{beck}.
An adaptive finite element code, using cubic polynomial shape functions~\cite{pask}, has a convergence rate of $h^6$.
Finite difference methods have sometimes low~\cite{gpaw} $h^3$ or high convergence 
rates~\cite{chelikowsky} but are not adaptive. 

As discussed above, localization in real space is essential for molecular systems. Basis sets that are not localized in real space are wasteful in this context. For instance, with plane waves one has to fill an orthorhombic cell into which the molecule fits. 
Large subregions of the cell may contain no atoms and therefore no charge density, 
but this feature can not be exploited with plane waves. 
Since Daubechies wavelets have a compact support, one can consistently define a set of localization parameters which allows us to put the basis functions only on the points which are sufficiently close to the atoms. 
The computational volume in our method is thus given only by the union of spheres centered on all the atoms in the system.
Real space localization is also necessary for the implementation of linear scaling algorithms~\cite{RMPSG}. This basis set is thus a promising candidate for developing such algorithms.

Localization in Fourier space is useful for preconditioning purposes. For a given system, the convergence rate of the minimization process depends on the highest eigenvalue of the Hamiltonian operator.
Since the high frequency spectrum of the Hamiltonian is dominated by the kinetic energy operator, high kinetic energy basis functions are therefore also approximate eigenfunctions of the Hamiltonian.
A function localized in Fourier space is an approximate eigenfunction of the kinetic energy operator. 
By using such functions as basis functions for the KS orbitals the high energy spectrum of the Hamiltonian can thus easily be preconditioned. 

A high degree of adaptivity is necessary for all-electron calculations since 
highly localized core electrons require a much higher spatial resolution than the 
valence wavefunction away from the atomic core. High adaptivity can in principle 
be obtained with a wavelet basis and wavelet based all-electron electronic 
structure programs have been developed~\cite{arias,harrison}. In contrast to these 
developments we use pseudopotentials since such pseudopotentials are the easiest 
way to incorporate the relativistic effects that are important for heavy elements.
The use of pseudopotentials drastically reduces the need for adaptivity and we 
have therefore only two levels of adaptivity. 
We have a high resolution region that contains all the chemical bonds and a low resolution region further away from the atoms where the wavefunctions decay exponentially to zero. In the low resolution region each grid point carries a single basis function. In the high resolution region it carries in addition 7 wavelets. In terms of degrees of freedom, the high resolution region is thus 8 times denser than the low resolution region. In comparison with a plane wave methods our wavelet method is therefore particularly efficient for open structures with large empty spaces and a relatively small bonding region. 

The outline of this paper is as follows: in the next section we describe the fundamental properties of Daubechies wavelets. Then we will describe how the various operations needed in an electronic structure 
calculations are done in a scaling function/wavelet basis. 
The last part of the paper illustrates the performances of our DFT code based on Daubechies wavelets.

\section{Adaptivity in a wavelet basis}
There are two fundamental functions in wavelet theory~\cite{daub,ppur}, 
the scaling function $\phi(x)$ and the wavelet $\psi(x)$. 

\begin{figure}[h]
\begin{center}
\includegraphics[width=.45\textwidth]{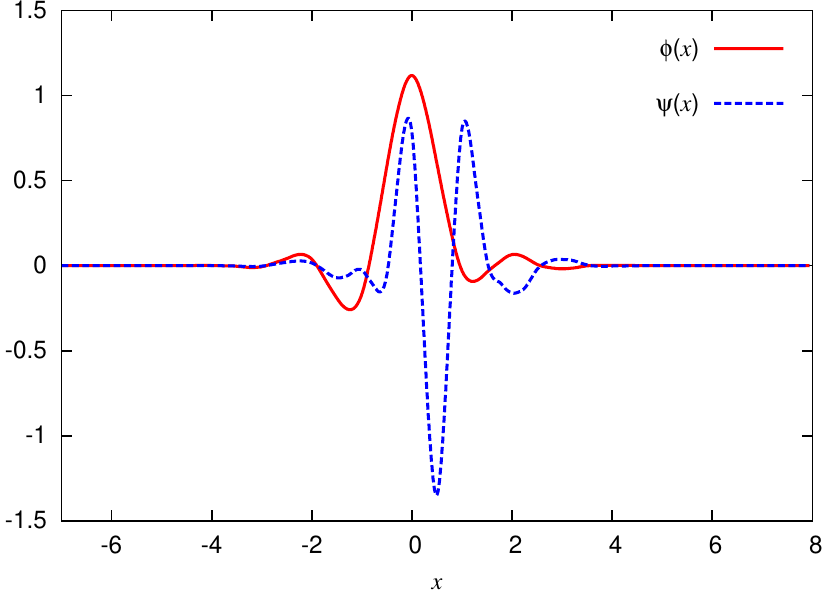}
\caption{ \label{D16} Daubechies scaling function $\phi$ and wavelet $\psi$ of order 16.
Both are different from zero only in the interval from -7 to 8.}
\end{center}
\end{figure}

The most important property of these functions is that they satisfy the so-called refinement equations
\begin{align}
\label{refinement}
\phi(x) &= \sqrt 2\sum_{j=1-m}^{m} \text{\sl h}_j \: \phi(2 x -j) \\
\psi(x) &= \sqrt 2\sum_{j=1-m}^{m} \text{\sl g}_j \: \phi(2 x -j) \notag
\end{align}
which establishes a relation between the scaling functions on a grid with grid spacing $h$ and another 
one with spacing $h/2$. 
$\text{\sl h}_j$ and $\text{\sl g}_j=(-1)^j \text{\sl h}_{-j+1}$ are the elements of a filter that 
characterizes the wavelet family, and $m$ is the order of the scaling function-wavelet family. 
All the properties of these functions can be obtained from the relations \eqref{refinement}.
The full basis set can be obtained from all translations 
by a certain grid spacing $h$ of the mother function centered at the origin.
The mother function is localized, with compact support. The maximally symmetric Daubechies scaling function and wavelet of order 16 that are used in this work are shown in Fig.~\ref{D16}.

For a three-dimensional description, the simplest basis set is obtained by a set of products of equally spaced scaling functions on a grid of grid spacing $h'$ 
\begin{equation} \label{bsst}
 \phi_{i,j,k}\left({\bf r}\right) = \phi(x/h'-i)\, \phi(y/h'-j)\, \phi(z/h'-k)\;. 
\end{equation}
In other terms, the three-dimensional basis functions are a tensor product of one dimensional basis functions.
Note that we are using a cubic grid, where the grid spacing is the same in all directions, but the following description can be straightforwardly applied to general orthorombic grids.

The basis set of Eq.~\ref{bsst} is equivalent to a mixed basis set of scaling functions on a twice coarser grid of grid spacing $h=2 h'$
\begin{equation}
\phi_{i,j,k}({\bf r}) = \phi(x/h-i)\, \phi(y/h-j)\, \phi(z/h-k) \label{scf}
\end{equation}
augmented by a set of 7 wavelets 
\begin{eqnarray}
 \psi^{1}_{i,j,k}({\bf r}) & = &  \psi(x/h-i)\, \phi(y/h-j)\, \phi(z/h-k)\nonumber  \\
 \psi^{2}_{i,j,k}({\bf r}) & = &  \phi(x/h-i)\, \psi(y/h-j)\, \phi(z/h-k)\nonumber  \\
 \psi^{3}_{i,j,k}({\bf r}) & = &  \psi(x/h-i)\, \psi(y/h-j)\, \phi(z/h-k)\nonumber  \\
 \psi^{4}_{i,j,k}({\bf r}) & = &  \phi(x/h-i)\, \phi(y/h-j)\, \psi(z/h-k) \label{wvl}\\
 \psi^{5}_{i,j,k}({\bf r}) & = &  \psi(x/h-i)\, \phi(y/h-j)\, \psi(z/h-k)\nonumber  \\
 \psi^{6}_{i,j,k}({\bf r}) & = &  \phi(x/h-i)\, \psi(y/h-j)\, \psi(z/h-k)\nonumber  \\
 \psi^{7}_{i,j,k}({\bf r}) & = &  \psi(x/h-i)\, \psi(y/h-j)\, \psi(z/h-k)\nonumber
\end{eqnarray}
This equivalence follows from the fact that, from Eq. \eqref{refinement}, every scaling function and wavelet on a coarse grid of spacing $h$ can be expressed as a linear combination of scaling functions at the fine grid level $h'$ and vice versa.

\begin{figure}[h]
\begin{center}
\includegraphics[width=.5\textwidth]{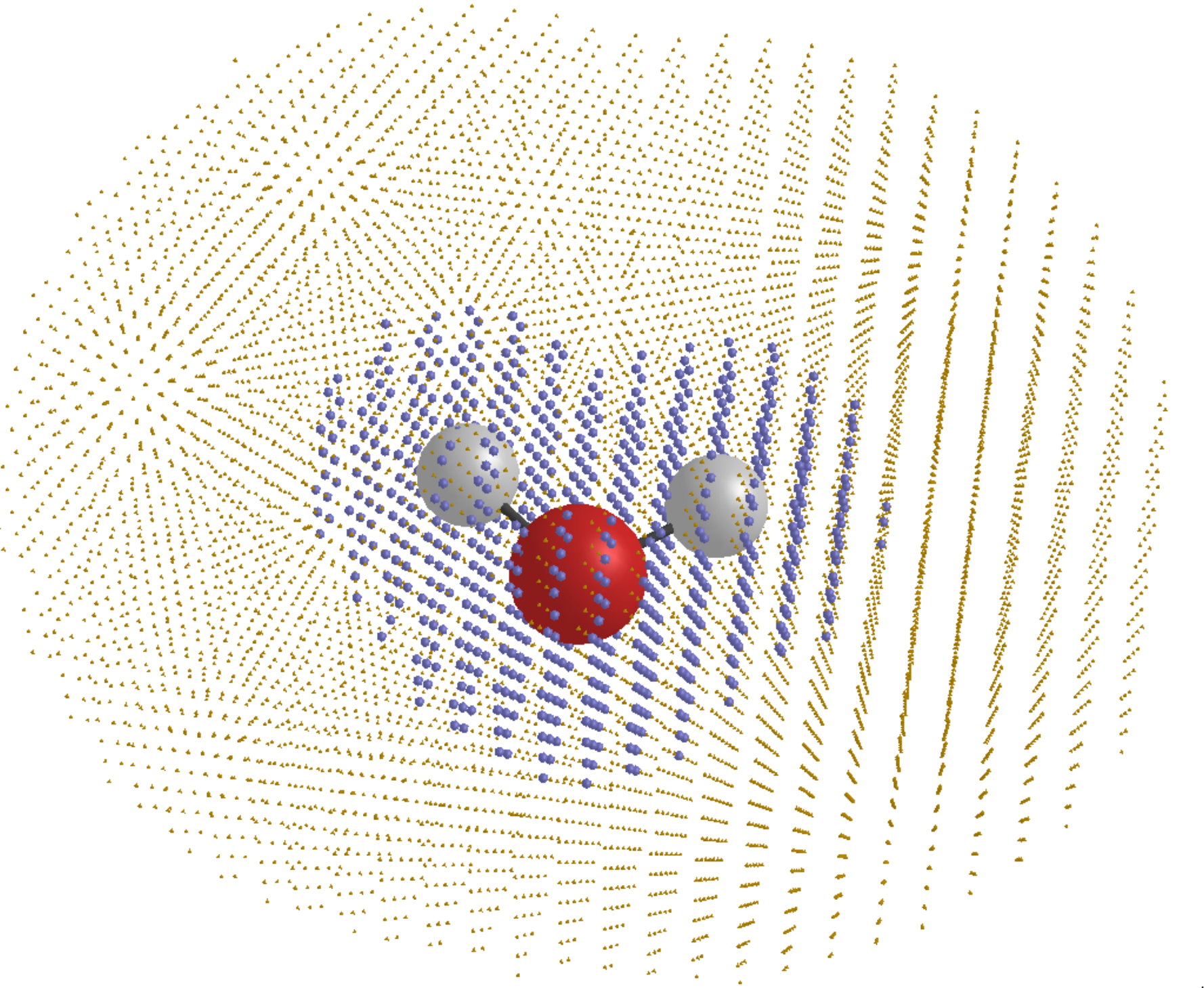}
\caption{ \label{grid} A 2-level adaptive grid around a H$_2$O molecule. 
The high resolution grid points carrying both scaling functions and wavelets are 
shown in blue (larger points), the low resolution grid points carrying only a single scaling function are shown in yellow (smaller points).}
\end{center}
\end{figure}

The points of the simulation grid fall into 3 different classes. The points which are 
very far from the atoms will have virtually zero charge density and thus will not carry any
basis functions. 
The remaining grid points are either in the high resolution region which contains the 
chemical bonds or in the low resolution regions which contains the exponentially decaying 
tails of the wavefunctions. 
In the low resolution region one uses only one scaling function per coarse grid point, 
whereas in the high resolution region one uses both the scaling function and the 7 wavelets. 
In this region the resolution is thus doubled in each spatial dimension compared to the low resolution region. 
Fig.~\ref{grid} shows the 2-level adaptive grid around a water molecule.

A wavefunction $\Psi(\mathbf r)$ can thus be expanded in this basis:
\begin{multline} \label{wavef}
\Psi({\bf r}) = \sum_{i_1,i_2,i_3} s_{i_1,i_2,i_3} \phi_{i_1,i_2,i_3}(\mathbf r) +\\
+ \sum_{j_1,j_2,j_3} \sum_{\nu=1}^7 d^\nu_{j_1,j_2,j_3} \psi^{\nu}_{j_1,j_2,j_3}(\mathbf r)  
\end{multline}
The sum over $i_1$, $i_2$, $i_3$ runs over all the grid points contained in the low resolution region and the sum over $j_1$, $j_2$, $j_3$ over all the points contained in the smaller high resolution region.

The decomposition of scaling function into coarser scaling functions and wavelets can 
be continued recursively to obtain more than 2 resolution levels. We found however that a high degree of adaptivity is not of paramount importance in pseudopotential calculations. In other terms, the pseudopotentials smooth the wavefunctions so that two levels of resolution are enough in most cases to achieve good computational accuracy. In addition, more than two resolution levels lead to more complicated algorithms such as the non-standard operator form~\cite{nonstand} that, in turn, lead to larger prefactors.

The transformation from a pure fine scaling function representation (a basis set which contains only scaling functions centered on a finer grid of spacing $h'$) to a mixed 
coarse scaling function/wavelet representation is done by the fast wavelet 
transformation~\cite{ppur} which is a convolution and scales linearly with 
respect to the number of basis functions being transformed.

The wavefunctions are stored in a compressed form where only the nonzero scaling function and wavelets coefficients are stored. The basis set being orthogonal, several operations such as scalar products among different orbitals and between orbitals and the projectors of the non-local pseudopotential can directly be done in this compressed form. 
In the following sections we will illustrate the main operations which must be performed in the 
context of a DFT calculation.

\section{Overview of the method}
In the KS formulation of DFT, the electronic density of a system of N electrons can be calculated from the square modulus of a set of wavefunctions:
\be
\rho(\mathbf r) = \sum_{i=1}^{N/2} n_\text{occ}^{(i)} \left|\Psi_i(\mathbf r)\right |^2\;,
\ee
where the KS wavefunctions $|\Psi_i\rangle$ are eigenfunctions of the KS Hamiltonian, with pseudopotential $V_\text{psp}$:
\be
\left(-\frac{1}{2} \nabla^2 + V_\text{KS}[\rho] + V_\text{psp} \right)|\Psi_i\rangle=\epsilon_i |\Psi_i\rangle\;.
\ee
For the sake of simplicity we assume in this description that our electronic system is a closed-shell system of non-spin-polarised electronic orbitals. For this reasons we have exactly $N/2$ KS wavefunctions and $ \forall i\;\; n_\text{occ}^{(i)}=2 $.

The KS potential
\be
V_\text{KS}[\rho] = V_H[\rho] + V_\text{xc}[\rho] + V_\text{ext}\;,
\ee
contains the Hartree potential, solution of the Poisson's equation $\nabla^2 V_H = - 4 \pi \rho$, the exchange-correlation potential $V_\text{xc}$ and the external ionic potential $V_\text{ext}$ acting on the electrons. The method we illustrate in this paper is conceived for isolated systems, namely free boundary conditions.

In our method, we choose the pseudopotential term $V_\text{psp}$ to be of the form of norm-conserving GTH-HGH pseudopotentials~\cite{gth,hgh,krack}, which have a local and a nonlocal term, $V_\text{psp}=V_\text{local} + V_\text{nonlocal}$. For each of the ions these potentials have this form:
\begin{align}
 V_\text{local}(\mathbf r)&= -\frac{Z_{ion}}{r} \text{\rm erf}\left(\frac{r}{\sqrt 2 r_\text{loc}} \right) +
\exp\left[-\frac{1}{2}\left(\frac{r}{r_\text{loc}} \right)^2\right] \times  \notag \\
&\times \left[C_1 + C_2 \left(\frac{r}{r_\text{loc}} \right)^2 + C_3 \left(\frac{r}{r_\text{loc}} \right)^4 + C_4\left(\frac{r}{r_\text{loc}} \right)^6 \right]  \\
V_\text{nonlocal} &= \sum_\ell \sum_{i,j=1}^3 h_{ij}^{(\ell)} |p_i^{(\ell)}\rangle \langle p_j^{(\ell)}| \\
\langle \mathbf r | p_i^{(\ell)} \rangle &= \frac{\sqrt 2 r^{\ell+2(i-1)} \exp\left[-\frac{1}{2}\left(\frac{r}{r_\ell} \right)^2\right]}{r_\ell^{\ell + (4 i -1)/2} \sqrt{\Gamma\left(\ell + \frac{4 i-1}{2}\right)}} \sum_{m=-\ell}^{+\ell}Y_{\ell m}(\theta,\phi) \notag\;,
\end{align}
where $Y_{\ell m}$ are the spherical harmonics, and $r_\text{loc}$, $r_\ell$ are, respectively, the localization radius of  the local pseudopotential term and of each projector.

The analytic form of the pseudopotentials together with the fact that their expression in real space can be written in terms of a linear combination of tensor products of one dimensional functions is of great utility in our method.

Each term in the Hamiltonian is implemented differently, and will be illustrated in the following sections. After the application of the Hamiltonian, the KS wavefunctions are updated via a direct minimisation scheme~\cite{payne}, which in its actual implementation is fast and reliable for non-zero gap systems, namely insulators. At present we have concentrated on systems with a gap, however we see no reason why the method can not be extended to metallic systems.

\section{Treatment of kinetic energy}
The matrix elements of the kinetic energy operator among the basis functions 
of our mixed representation (i.e. scaling functions with scaling functions, 
scaling function with wavelets and wavelets with wavelets) can be calculated 
analytically~\cite{beylkin}. For simplicity, let us illustrate the 
application of the kinetic energy operator onto 
a wavefunction  $\Psi$ that is only expressed in terms of scaling functions. 
$$ \Psi(x,y,z) = \sum_{i_1,i_2,i_3}\!\!\! s_{i_1,i_2,i_3} 
    \phi(x/h-i_1)\, \phi(y/h-i_2)\, \phi(z/h-i_3) $$
The result of the application of the kinetic energy operator on this wavefunction, projected to the original scaling function space, has the expansion coefficients 
\begin{eqnarray}
\hat{s}_{i_1,i_2,i_3}=-\frac{1}{2h^3}\int  \phi(x/h-i_1)\, \phi(y/h-i_2)\, \phi(z/h-i_3)\times\nonumber\\
\times\Delta \Psi(x,y,z){\rm dx dy dz}
\nonumber
\end{eqnarray}

Analytically the coefficients $s_{i_1,i_2,i_3}$ and $\hat{s}_{i_1,i_2,i_3}$ are related by a convolution
\begin{equation} \label{kiner}
 \hat{s}_{i_1,i_2,i_3} =\frac{1}{2} \sum_{j_1,j_2,j_3} K_{i_1-j_1,i_2-j_2,i_3-j_3} s_{j_1,j_2,j_3} 
\end{equation}
where 
\begin{equation} \label{fprod}
K_{i_1,i_2,i_3}=T_{i_1} T_{i_2} T_{i_3},
\end{equation}
and
\be
 T_{i_1}=  \int \dd x\, \phi(x/h-i_1) \: \partial_x^2 \phi(x/h)\;.
\ee
Using the refinement equation \eqref{refinement}, the values of the $T_i$ can be calculated analytically, from a suitable eigenvector of a matrix derived from the wavelet filters~\cite{beylkin}. For this reason the expression of the kinetic energy operator is \emph{exact} in a given Daubechies basis.

Since the 3-dimensional kinetic energy filter $K_{i_1,i_2,i_3}$
is a product of three one-dimensional filters (Eq.~\ref{fprod}) the convolution in Eq.~\ref{kiner} can be evaluated with $3 N_1 N_2 N_3 L$ operations for a three-dimensional grid of $ N_1 N_2 N_3 $ grid points.
$L$ is the length of the one-dimensional filter which is 29 for our Daubechies family. 
The kinetic energy can thus be evaluated with linear scaling with respect to the 
number of nonvanishing expansion coefficients of the wavefunction. 
This statement remains true for a mixed scaling function-wavelet basis where we have both nonvanishing $s$ and $d$ coefficients and 
for the case where the low and high resolution regions cover only parts 
of the cube of $N_1 N_2 N_3$ grid points. 

\begin{figure}[h]
\begin{center}
\includegraphics[width=.45\textwidth]{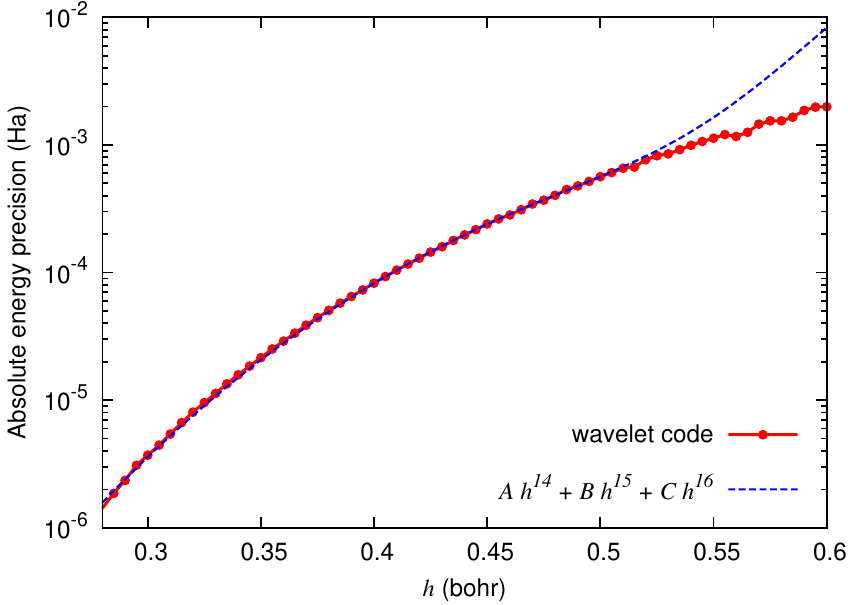}
\caption{Convergence rate $\mathcal O (h^{14})$ of the wavelet code for a test run on a carbon atom. For this run the interpolation parameters are found to be, within 2\% accuracy: $A=344$, $B=-1239$, $C=1139$. Other test systems gave comparable convergence rates.}
\label{convrate}
\end{center}
\end{figure}

The Daubechies wavefunctions of degree 16 have an approximation error of $h^{8}$, i.e. 
the difference between the exact wavefunction and its representation in a finite basis 
set (Eq.~\ref{wavef}) is decreasing as $h^{8}$. The error of the kinetic 
energy in a variational scheme decreases then as $h^{2\cdot 8 -2}=h^{14}$~\cite{finel}. 
As we will see the kinetic energy is limiting the convergence rate in our scheme and 
the overall convergence rate is thus $h^{14}$. 
Figure \ref{convrate} shows this asymptotic convergence rate.

\section{Treatment of local potential energy} \label{potentialsection}
In spite of the striking advantages of Daubechies wavelets the initial exploration
of this basis set~\cite{wang} did not lead to any algorithm that would be useful for
real electronic structure calculations. This was due to the fact that an accurate evaluation
of the local potential energy is difficult in a Daubechies wavelet basis.

By definition, the local potential $V(\mathbf r)$ can be easily known
on the nodes of the uniform grid of the simulation box.
Approximating a potential energy matrix element $V_{i,j,k;i',j',k'}$
$$ V_{i,j,k;i',j',k'} =
\int \dd \mathbf{r} \phi_{i',j',k'}({\bf r})  V({\bf r}) \phi_{i,j,k}({\bf r}) $$
by
$$ V_{i,j,k;i',j',k'} \approx \sum_{l,m,n}  \phi_{i',j',k'}({\bf r}_{l,m,n})  V({\bf r}_{l,m,n}) \phi_{i,j,k}({\bf
r}_{l,m,n}) $$
gives an extremely slow convergence rate with respect to the number of grid points
used to approximate the integral because a single scaling function is not very smooth, 
i.e. it has a rather low number of continuous derivatives. 
A. Neelov~ and S. Goedecker~\cite{magic} have shown that one should not 
try to approximate a single matrix element as accurately as possible
but that one should try instead to approximate directly the
expectation value of the local potential. The reason for this strategy
is that the wavefunction expressed in the Daubechy basis is smoother than a single Daubechies basis function.
A single Daubechies scaling function of order 16 has only 4 continuous derivatives.
By suitable linear combinations of Daubechies 16 one can however exactly represent
polynomials up to degree 7, i.e functions that have 7 non-vanishing continuous derivatives.
The discontinuities get thus canceled by taking suitable linear combinations.
Since we use pseudopotentials, our exact wavefunctions are analytic and can
locally be represented by a Taylor series. We are thus approximating functions
that are approximately polynomials of order 7 and the discontinuities nearly cancel.

\begin{figure}[h]
\begin{center}
\includegraphics[width=.45\textwidth]{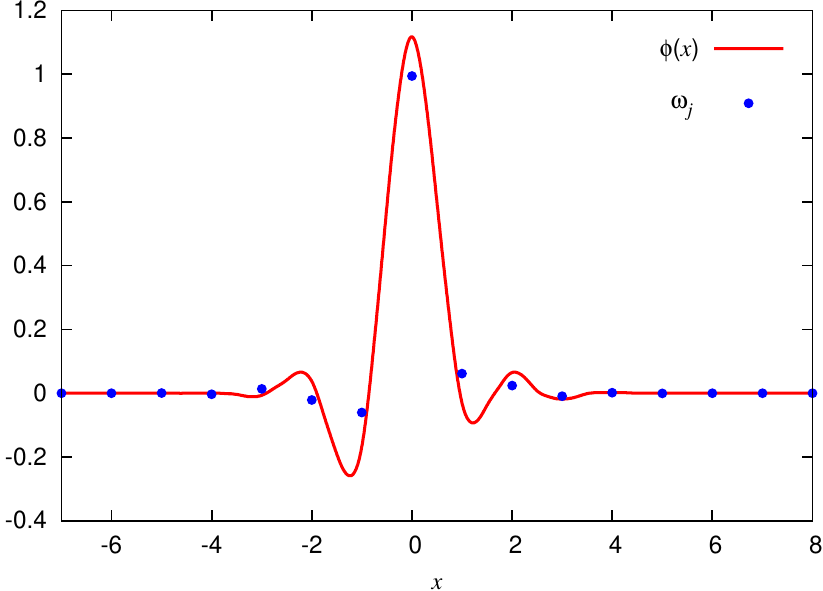}
\caption{The magic filter $\omega_i$ for the least asymmetric Daubechies-16 basis.}
\label{fig:magic}
\end{center}
\end{figure}

Instead of calculating the exact matrix elements we therefore use
matrix elements with respect to a smoothed version $\tilde{\phi}$ of the
Daubechies scaling functions.
\begin{multline} \label{vmag}
 V_{i,j,k;i',j',k'} \approx  \sum_{l,m,n}  \tilde{\phi}_{i',j',k'}({\bf r}_{l,m,n})  V({\bf r}_{l,m,n}) \tilde{\phi}_{i,j,k}({\bf r}_{l,m,n}) =\\
  \sum_{l,m,n}  \tilde{\phi}_{0,0,0}({\bf r}_{l-i',m-j',n-k'})  V({\bf r}_{l,m,n}) \tilde{\phi}_{0,0,0}({\bf r}_{l-i,m-j,n-k}) 
\end{multline}
where the smoothed wave function is defined by
$$ \tilde{\phi}_{0,0,0}({\bf r}_{l,m,n})=\omega_l\omega_m\omega_n$$
and $\omega_l$ is the ``magic filter''.
The relation between the true functional values, i.e. the scaling function, and $\omega$ is shown in figure~\ref{fig:magic}.
Even though Eq.~\ref{vmag} is not a particulary good approximation for a single matrix
element it gives an excellent approximation for the expectation values of the local potential energy
$$ \int dx \int dy  \int dz \Psi(x,y,z)  V(x,y,z) \Psi(x,y,z) $$
and also for matrix elements between different wavefunctions 
$$ \int dx \int dy  \int dz \Psi_i(x,y,z)  V(x,y,z) \Psi_j(x,y,z) $$
in case they are needed.
In practice we do not explicitly calculate any matrix elements but we
apply only filters to the wavefunction expansion coefficients as will be shown
in the following. This is mathematically equivalent but numerically much more
efficient.

\begin{figure}[h]
\begin{center}
\includegraphics[width=.45\textwidth]{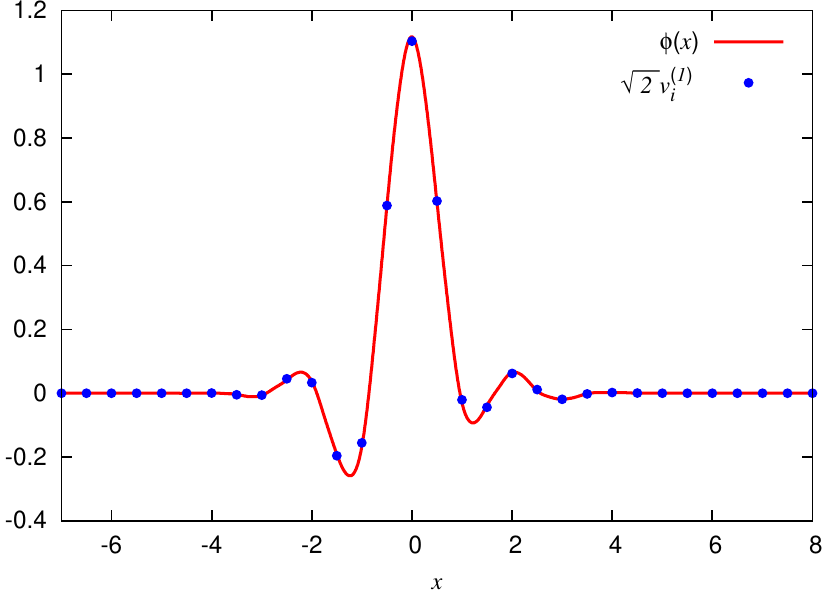}
\caption{The fine scale magic filter $v_{i}^{(1)}$ (combination of a wavelet transform and the magic filter in figure \ref{fig:magic}) for the least asymmetric Daubechies-16 basis, scaled by $\sqrt{2}$ for comparison with the scaling function.
The values of the filter on the graph are almost undistinguishable from the values of the scaling function. However, there is a slight difference which is important for the correct asymptotic convergence at small values of grid spacing $h$.}
\label{fig:magic2}
\end{center}
\end{figure}

\begin{figure}[h]
\begin{center}
\includegraphics[width=.45\textwidth]{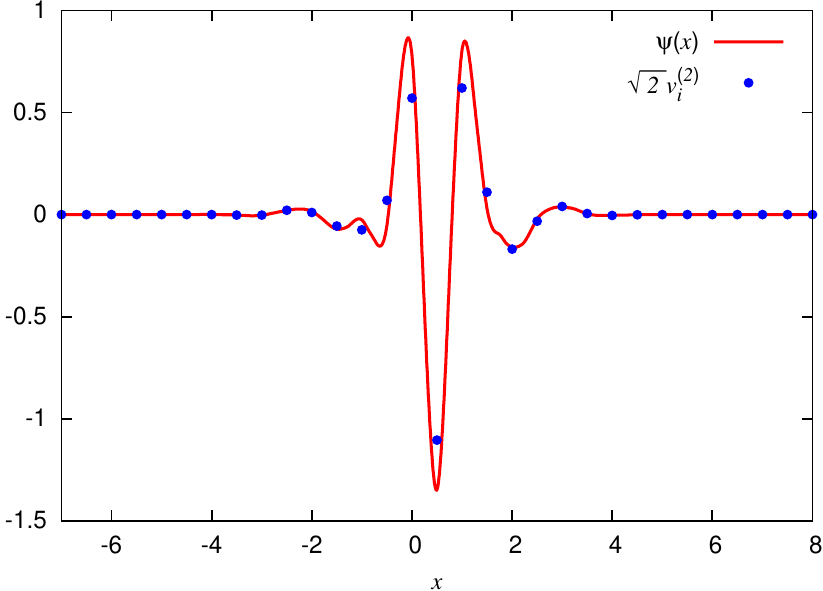}
\caption{The fine scale magic filter $v_{i}^{(2)}$ (combination of a wavelet transform and the magic filter in figure \ref{fig:magic}) for the least asymmetric Daubechies-16 {\it wavelet}, scaled by $\sqrt{2}$ for comparison with the wavelet itself.}
\label{fig:magic3}
\end{center}
\end{figure}

Since the operations with the local potential $V$ are performed in the computational box on the double resolution grid with grid spacing $h'=h/2$, we must perform a wavelet transformation before applying the magic filters. These two operations can be combined in one, giving rise to modified magic filters both for scaling functions and wavelets on the original grid of spacing $h$. These modified magic filters can be obtained from the original ones using the refinement relations and they are shown in Figures~\ref{fig:magic2} and \ref{fig:magic3}. 
Following the same guidelines as the kinetic energy filters, the smoothed real space values $\tilde{\Psi}_{i,j,k}$ of a wavefunction $\Psi$ are calculated by performing a product of three one-dimensional convolutions with the magic filters along the $x$, $y$ and $z$ directions. 
For the scaling function part of the wavefunction the corresponding formula is 
$$\tilde{\Psi}_{i_1,i_2,i_3} = 
   \sum_{j_1,j_2,j_3} s_{j_1,j_2,j_3}  v_{i_1-2 j_1}^{(1)} v^{(1)}_{i_2-2 j_2} v^{(1)}_{i_3-2 j_3} $$
where $v_{i}^{(1)}$ is the filter that maps a scaling function on a double resolution 
grid. Similar convolutions are needed for the wavelet part. 
The calculation is thus similar to the treatment of the Laplacian in the kinetic energy. 

Once we have calculated $\tilde{\Psi}_{i,j,k}$ the approximate expectation value $\epsilon_V$ of the local potential $V$ for a wavefunction $\Psi$ is obtained by simple summation on the double resolution real space grid:
$$ \epsilon_V = \sum_{j_1,j_2,j_3} 
                \tilde{\Psi}_{j_1,j_2,j_3}  V_{j_1,j_2,j_3} \tilde{\Psi}_{j_1,j_2,j_3} $$

The evaluation of the local potential energy $\epsilon_V$ converges with a convergence rate of $h^{16}$ to the exact value where $h$ is the grid spacing. 
Therefore, the potential energy has a convergence rate two powers of $h$ faster than the rate for the kinetic energy. 

\section{Calculation of Hartree potential}
We saw in the section on the treatment of the local potential energy how to express efficiently the point values of the smoothed wavefunction $\tilde{\Psi}$ on the fine grid mesh.
From these values the charge density on a grid point $j_1,j_2,j_3$ of the double resolution grid is given by
\begin{equation} \label{sumrho}
\rho_{j_1,j_2,j_3} = \sum_{i} n^{(i)}_\text{occ} \tilde{\Psi}_{i;j_1,j_2,j_3}^2 
\end{equation}
where $n^{(i)}_\text{occ}$ are the occupation numbers. For a closed shell system they 
equal 2 for the occupied orbitals and zero for all other orbitals. 
The discrete charge density $\rho_{j_1,j_2,j_3}$ is a very good approximation to the charge distribution of the continuous wavefunctions $|\Psi\rangle $ in the sense that the first multipoles of the discrete charge distribution converge rapidly to the values of the continuous charge distribution.
The monopole converges with a rate of $h^{16}$. For each higher multipole moment the 
convergence rate is reduced by one power of $h$, i.e. dipoles converge with a rate of $h^{15}$, quadrupoles with $h^{14}$, etc. 
The discrete charge density $\rho$ on the double resolution grid is then the input 
to various Poisson solvers that are available for different boundary conditions. 
In the case of free boundary conditions, appropriate for isolated molecules, the values $\rho_{j_1,j_2,j_3}$ form the coefficients for an expansion in interpolating scaling functions of order 16.  This expansion strictly conserves all the multipoles up to the angular moment $\ell=15$ and allows to solve the integral equation for the potential explicitly with the correct boundary conditions~\cite{PSfreeBC}.
In addition to free boundary conditions we have also implemented surface boundary 
conditions~\cite{PSsurfacesBC}, i.e. periodicity in 2 directions and free boundary conditions in the third direction. In this case the charge density is represented in a mixed plane wave-scaling function representation. 

These Poisson solvers have a convergence rate of ${h'}^m$, where $m$ is the order of the 
interpolating scaling functions used to express the Poisson kernel. Since we use interpolating scaling functions of order 16 the convergence rate of the electrostatic potential is faster than the rate for the kinetic energy.
All these Poisson Solvers have one thing in common, they perform explicitly the convolution of the density with the Green's functions of the Poisson's equation. 
The necessary convolutions are done by a traditional zero-padded FFT procedure which leads to an $\mathcal O(N \log N)$ operation count with respect to the number of grid points $N$.
The accuracy of the potential is uniform over the whole volume and one can thus use the smallest possible volume compatible with the requirement that the tails of the wavefunctions have decayed to very small values at the surface of this volume.
The fraction of the computational time needed for the solution of the Poisson's equation 
decreases with increasing system size and is roughly 1\% for large systems, see section \ref{performances}. 
Moreover, the explicit Green's function treatment of the Poisson's solver allows us to treat isolated systems with a net charge directly without the insertion of compensating charges. 

\section{XC functionals and implementation of GGA's}
The charge density expression used to calculate the Hartree potential 
\be
\rho(\mathbf r)= \sum_{i} n^{(i)}_\text{occ} |\tilde{\Psi}_{i}(\mathbf r)|^2\;,
\ee
is also used for the calculation of the exchange correlation energy $E_{\text{xc}}$ 
and the corresponding potential $V_{\text{xc}}$.
Any real-space based implementation of the XC functionals fits well with this density 
representation. In our program we use the XC functionals as implemented in ABINIT code. 
To this aim, we use the same ABINIT XC routines to calculate the exchange correlation energy
\be
E_{\text{xc}}=\int \rho(\mathbf r) \epsilon_{\text{xc}} (\mathbf r) \dd \mathbf r\;,
\ee
together with the XC potential 
\be
V_{\text{xc}}(\mathbf r)=\frac{\delta E_{\text{xc}}}{\delta \rho(\mathbf r)}\;.
\ee
Also spin-polarised (collinear) version of the ABINIT XC functionals can be used with our method. 

In the case of GGA exchange-correlation functionals the XC energy density depends both on 
the local values of the charge density $\rho$ and on the modulus of its gradient:
\be
\epsilon_{\text{xc}} (\mathbf r)=\epsilon_{\text{xc}} \left(\rho(\mathbf r),|\mathbf{\nabla} \rho| (\mathbf r)\right)\;.
\ee
A traditional finite difference scheme of fourth order is used on the double resolution 
grid to calculate the gradient of the charge density
\be\label{gradient}
\partial_w \rho({ \bf r}_{i_1,i_2,i_3})=\sum_{j_1,j_2,j_3} c_{i_1,i_2,i_3;j_1,j_2,j_3}^{(t)} 
   \rho_{j_1,j_2,j_3} \;,
\ee
where $w=x,y,z$. For grid points close to the boundary of the computational volume
the above formula requires grid points outside the volume. 
For free boundary conditions the values of the charge density outside the computational volume in a given direction are taken to be equal to the value at the border of the grid. 

The relation between the gradient and the density must be taken into account when calculating 
$V_{\text{xc}}$ in the standard White-Bird approach~\cite{whitebird}, where the density gradient is considered as an explicit functional of the density. 
There the XC potential can be split in two terms:
\begin{align}
 V_{\text{xc}}(\mathbf{r}_{i_1,i_2,i_3})&=
V_{\text{xc}}^{o}(\mathbf r)+V_{\text{xc}}^{c}(\mathbf r)\;,\notag\\
\end{align}
where
\begin{align}
V_{\text{xc}}^{o}(\mathbf{r}_{i_1,i_2,i_3})&=
\epsilon_{\text{xc}} (\mathbf r) +\rho(\mathbf r) \frac{\d \epsilon_{\text{xc}}}{\d \rho}(\mathbf r)\;,\\
V_{\text{xc}}^{c}(\mathbf{r}_{i_1,i_2,i_3})&=
\sum_{j_1,j_2,j_3}\frac{\rho}{|\nabla \rho|} \frac{\d \epsilon_{\text{xc}}}{\d |\nabla \rho|}(\mathbf{r}_{j_1,j_2,j_3}) \times\notag\\ &\times \sum_{w=x,y,z} \d_w \rho (\mathbf{r}_{j_1,j_2,j_3}) c_{j_1,j_2,j_3;i_1,i_2,i_3}^{(w)}\;,\notag
\end{align}
where the ``ordinary'' part $V_{\text{xc}}^{o}$ is present in the same form of LDA functionals, while the White-Bird ``correction'' term $V_{\text{xc}}^{c}$ appears only when
the XC energy depends explicitly on $|\nabla \rho|$.
The $c^{(w)}$ are the coefficients of the finite difference formula used to calculate the gradient of the charge density~\eqref{gradient}. 

The evaluation of the XC terms and also, when needed, the calculation of the gradient of the charge density, may easily be performed together with the Poisson solver used to evaluate the Hartree potential.  This allows us to save computational time. 

\section{Treatment of the non-local pseudopotential}\label{nonlocalpseudosection}
The energy contributions from the non-local pseudopotential have for each angular 
moment $l$ the form
$$\sum_{i,j} \langle \Psi | p_i \rangle h_{ij} \langle p_j | \Psi \rangle $$
where $| p_i \rangle$ is a pseudopotential projector. 
Once applying the hamiltonian operator, the application of one projector on the wavefunctions  requires the calculation of 
$$  |\Psi\rangle \rightarrow |\Psi \rangle + \sum_{i,j} |p_i \rangle h_{ij}  \langle p_j | \Psi \rangle\;. $$
If we use for the projectors the representation of Eq.~\ref{wavef} (i.e. the same as for 
the wavefunctions) both operations are trivial to perform. Because of the orthogonality of the basis set we just have to calculate scalar products among the coefficient vectors and to update the wavefunctions. 
The scaling function and wavelet expansion coefficients for 
the projectors are given by~\cite{ppur}
\begin{align}
\int p(\mathbf r)\,& \phi_{i_1,i_2,i_3}({\bf r}) \dd{\bf r}\;, &
\int p(\mathbf r)\,& \psi^{\nu}_{i_1,i_2,i_3}({\bf r}) \dd{\bf r}\;.
\end{align}
where we used the notation (\ref{scf}),(\ref{wvl}).

The GTH-HGH pseudopotentials~\cite{gth,hgh} have projectors which are written in terms of gaussians times polynomials. This form of projectors is particularly convenient to be expanded in the Daubechies basis.
In other terms, since the general form of the projector is
$$\langle \mathbf r |p \rangle = e^{- c r^2}  x^{\ell_x}y^{\ell_y}z^{\ell_z}\;, $$
the 3-dimensional integrals can be calculated easily since they can be factorized into a product of 3 one-dimensional integrals.
\begin{align}
  \int \langle \mathbf r |p \rangle  \phi_{i_1,i_2,i_3}({\bf r}) \dd{\bf r} &=
W_{i_1}(c,\ell_x) W_{i_2}(c,\ell_y) W_{i_3}(c,\ell_x)\;,\\
W_{j}(c,\ell) &= \int_{-\infty}^{+\infty} e^{- c t^2} t^\ell \phi(t/h-j) \dd t
\end{align}

The one-dimensional integrals are calculated in the following way. We first calculate the 
scaling function expansion coefficients for scaling functions on a one-dimensional grid that is 16 times 
denser. The integration on this dense grid is done by 
the well-known quadrature introduced in \cite{John}, that coincides with the magic filter \cite{magic}. This integration scheme based on the magic filter has a convergence rate of $h^{16}$ and we gain therefore a factor of $16^{16}$ in accuracy by going to a denser grid. This means that the expansion coefficients are for reasonable grid spacings $h$ accurate to machine precision. After having obtained the expansion coefficients with respect to the fine scaling functions we obtain the expansion coefficients with respect to the scaling functions and wavelets on the required resolution level by one-dimensional fast wavelet transformations. 
No accuracy is lost in the wavelet transforms and our representation of the projectors is therefore typically accurate to nearly machine precision. 

\section{Preconditioning}
As already mentioned, direct minimisation of the total energy is used to find the 
converged wavefunctions.
The gradient $g_i$ of the total energy with respect to the $i$-th wavefunction $|\Psi_i\rangle$ is given by 
\begin{equation} \label{gradef}
 | g_i \rangle = H | \Psi_i\rangle - \sum_j \Lambda_{ij} |\Psi_j \rangle \;,
\end{equation}
where $\Lambda_{ij} = \langle \psi_j | H | \psi_i\rangle$ are the Lagrange multipliers 
enforcing the orthogonality constraints.
Convergence is achieved when the average norm of the residue $\langle\overline{ g_i | g_i } \rangle^{1/2}$ is below an user-defined numerical tolerance.

Given the gradient direction at each step, several algorithms can be used to improve convergence. In our method we use either preconditioned steepest-descent algorithm or preconditioned DIIS method~\cite{pulay,hutter}.
These methods work very well to improve the convergence for non-zero gap systems if a good preconditioner is available. 

The preconditioning gradient $|\tilde g_i \rangle$  which approximately points in the direction of the minimum is obtained by solving the linear system of equations obtained by discretizing the equation
\begin{equation} \label{precon}
 \left( \frac{1}{2} \nabla^2 - \epsilon_i \right) \tilde{g}_i({\bf r}) = g_i({\bf r}) \;.
\end{equation}
The values $\epsilon_i$ are approximate eigenvalues obtained by a subspace 
diagonalization in a minimal basis of atomic pseudopotential orbitals  
during the generation of the input guess. For isolated systems, the values of the $\epsilon_i$ for the occupied states are always negative, therefore the operator of Eq. \eqref{precon} is positive definite.

Eq.~\eqref{precon} is solved by a preconditioned conjugate gradient (CG) method. The preconditioning is done by using the diagonal elements of the matrix representing the operator $\frac{1}{2}\nabla^2 - \epsilon_i$ in a scaling function-wavelet basis. In the initial step 
we use $\ell$ resolution levels of wavelets where $\ell$ is typically 4. 
To do this we have to enlarge the domain where the scaling function part of the 
gradient is defined to a grid that is a multiple of $2^\ell$. This means that the 
preconditioned gradient $\tilde{g}_i$ will also exist in a domain that is larger than 
the domain of the wavefunction $\Psi_i$. Nevertheless this approach is useful since it 
allows us to obtain rapidly a preconditioned gradient that has the correct overall 
shape. In the following iterations of the conjugate gradient we use only one wavelet 
level in addition to the scaling functions for preconditioning. In this way we can do 
the preconditioning exactly in the domain of basis functions that are used to 
represent the wavefunctions (Eq.~\ref{wavef}). A typical number of CG iterations necessary to obtain a meaningful preconditioned gradient is 5. 

\section{Orthogonalization}
We saw the need of keeping the wavefunctions $\Psi_i$ orthonormal at each step of the minimisation loop. 
This means that the overlap matrix $S$, with matrix elements
\begin{equation} \label{overlap}
S_{ij}= \langle \Psi_j | \Psi_i \rangle
\end{equation}
must be equal to the identity matrix. 

All the orthogonalization algorithms have a cubic complexity causing this part of the 
program to dominate for large systems, see Fig. \ref{percent}. We therefore optimized this part carefully and found that a pseudo-Gram-Schmidt algorithm that uses a Cholesky factorization of the overlap matrix $S$ is the most efficient method on parallel computers. In the following, we discuss the reasons for this choice by comparing it to two other orthogonalization algorithms: classical Gram-Schmidt and Loewdin orthogonalizations.

\subsection{Gram-Schmidt orthogonalization}
The classical Gram-Schmidt orthonormalization algorithm generates an orthogonal 
set of orbital $\left\{ |\overline{\Psi}_i\rangle \right\}$ out of a non-orthogonal set $\left\{ |\Psi_i \rangle \right\}$, by processing separately each orbital. The overlap of the currently processed orbital $|\Psi_i\rangle$ with the set of the already processed orbitals $\left\{|\overline{\Psi}_j \rangle \right\}_{j=1,\cdots, i-1}$ is calculated and is removed from $|\Psi_i\rangle$. Thereafter, the transformed orbital $|\overline{\Psi}_i\rangle$ is normalized.
\begin{equation} \label{gram-schmidt}
|\overline{\Psi}_i \rangle = | \Psi_i\rangle -\sum_{j=1}^{i-1} \langle \overline{\Psi}_j | \Psi_i \rangle | \overline{\Psi}_j \rangle
\end{equation}
\begin{equation}
| \overline{\Psi}_j \rangle \longrightarrow 
\frac{| \overline{\Psi}_j \rangle}{\sqrt{\langle \overline{\Psi}_j | \overline{\Psi}_j \rangle}}
\end{equation}
The algorithm consists of the calculation of $n(n+1)/2$  scalar products and 
wavefunction updates. If the coefficients of each orbital are distributed among 
several processors $n(n+1)/2$ communication steps are needed to sum up the various 
contributions from each processor to each scalar product. Such a large number 
of communication steps leads to a large latency overhead on a parallel computer 
and therefore to poor performances. 

\subsection{Loewdin orthogonalization}
The Loewdin orthonormalization algorithm is based on the following equation:
\begin{equation} \label{loewdin}
|\overline{\Psi}_i \rangle =\sum_j  S_{ij}^{-\frac{1}{2}} \: | \Psi_j \rangle \;,
\end{equation}
where a new set of orthonormal orbitals $|\overline{\Psi}_i \rangle$ is obtained by multiplying the inverse square-root of the overlap 
matrix $S$ with the original orbital set.

The implementation of this algorithm requires that the overlap matrix $S$ is calculated. 
As $S$ is a symmetric matrix, we need to calculate only a triangle of the original matrix which results in $n(n+1)/2$ scalar products. In contrast to the classical Gram-Schmidt algorithm the matrix elements $S_{ij}$ depend on the original set of orbitals and can be calculated in parallel in the case where each processor holds a certain subset of the coefficients of each wavefunction. At the end of this calculation a single communication step is needed to sum up the entire overlap matrix out of the contributions to each matrix element calculated by the different processors. 
Thereafter, the inverse square-root of $S$ is calculated. For this, we use the fact that $S$ is an hermitian positive definite matrix. Thus, there exist a unitary matrix $U$ which diagonalizes $S=U^\star\Lambda U$, 
where $\Lambda$ is a diagonal matrix with positive eigenvalues. Consequently, $S^{-\frac{1}{2}}=U^\dagger\Lambda^{-\frac{1}{2}} U$. 
Hence, an eigenvalue problem must be solved in order to find $U$ and $\Lambda$. 

\subsection{Pseudo Gram-Schmidt using Cholesky Factorization}
In this scheme a Cholesky factorization of the overlap matrix $S=LL^T$ is calculated. The new orthonormal orbitals are obtained by
\begin{equation} \label{cholesky}
| \overline{\Psi}_i \rangle=\sum_j\left(L_{ij}^{-1}\right)|\: \Psi_j \rangle \;,
\end{equation}
and are equivalent to the orbitals obtained by the classical Gram-Schmidt.
The procedure for calculating the overlap matrix out of the contributions 
calculated by each processor is identical to the Loewdin case.
Instead of solving an eigenvalue problem we have however to calculate the decomposition of the overlap matrix. This can be done much faster.
Thus, this algorithm has a lower pre-factor than the Loewdin scheme and requires only 
one communication step on a parallel computer.

\section{Calculation of forces}
Atomic forces can be calculated with the same method used for the application of 
the hamiltonian onto a wavefunction.
Since the scaling function/wavelet basis is not moving together with atoms, 
we have no Pulay forces~\cite{pulayforce} and atomic forces can be evaluated directly through the 
Feynman-Hellmann theorem.
Except for the force arising from the trivial ion-ion interaction, which for the $i$-th atom is
\be
\mathbf F_i^{(\text{ionic})} = \sum_{j \neq i} \frac{Z_i Z_j}{R_{ij}^3} (\mathbf R_i - \mathbf R_j)\;,
\ee
the energy terms which depend explicitly on the atom positions are related to the pseudopotentials.
As shown in the previous sections, the GTH-HGH pseudopotentials we are using are based on separable functions \cite{gth,hgh}, and can be splitted into a local and a non-local contribution.

For an atom $i$ placed at position $\mathbf R_i$, the contribution to the energy that comes from 
the local part of the pseudopotential is
\be
E_\text{local}(\mathbf R_i)=\int \dd \mathbf r \;V_\text{local}(|\mathbf r - \mathbf R_i|) \rho(\mathbf r)\;.
\ee
Where the local pseudopotential can be split into long and a short-ranged terms 
$V_\text{local}(\l)=V_L(\l)+V_S(\l)$, and
\begin{align}
V_L(\l) &= -\frac{Z_i}{\l} \text{erf}\left(\frac{\l}{\sqrt 2 r_\ell}\right) \notag \;,\\
V_S(\l) &= \exp\left(-\frac{\l^2}{2r_\ell^2}\right)\Biggl[C_1+ C_2 \left(\frac{\l}{r_\ell}\right)^2 + \\&+ C_3 \left(\frac{\l}{r_\ell}\right)^4 + C_4 \left(\frac{\l}{r_\ell}\right)^6\Biggr]\;,\notag
\end{align}
where the $C_i$ and $r_\ell$ are the pseudopotential parameters, depending on the atom of atomic number $Z_i$ under consideration. 
The energy contribution $E_\text{local}(\mathbf R_i)$ can be rewritten in an equivalent form. It is straightforward to verify that
\begin{multline} \label{ftrans}
E_\text{local}(\mathbf R_i)=\int \dd \mathbf r \;\rho_\text{L}(|\mathbf r - \mathbf R_i|) V_H(\mathbf r) \\+\int \dd \mathbf r  V_{S}(|\mathbf r - \mathbf R_i|) \rho(\mathbf r)\;,
\end{multline}
where $V_H$ is the Hartree potential, and $\rho_L$ is such that $\nabla_{\mathbf r}^2 V_L(|\mathbf r - \mathbf R_i|) = -4 \pi \rho_L(|\mathbf r - \mathbf R_i|)$. 
This analytical transformation remains also valid in our procedure for solving the 
discretized Poisson's equation. 
From equation ~\eqref{ftrans} we can calculate
\be
\rho_L (\l)=-\frac{1}{(2 \pi)^{3/2}}\frac{Z_i}{r_\ell^3} e^{-\frac{\l^2}{2r_\ell^2}}\;,
\ee
which is a localized (thus short-ranged) function.
The forces coming from the local pseudopotential are thus
\begin{multline}
\mathbf {F}_i^{(\text{local})} = - \frac{\d E_\ell(\mathbf R_i)}{\d \mathbf{R}_i} \\
 = \frac{1}{r_\ell}\int \dd \mathbf r \frac{\mathbf r -\mathbf{R_i}}{|\mathbf r - \mathbf R_i|} \Biggl[
 \;\rho'_L(|\mathbf r - \mathbf R_i|) V_H(\mathbf r) \\+ V'_S(|\mathbf r - \mathbf R_i|) \rho(\mathbf r)\Biggr]\;,
\end{multline}
where 
\begin{align}
\rho'_L(\l) &= \frac{1}{(2 \pi)^{3/2}}\frac{Z_{\text ion}}{r_{\text loc}^4} \l e^{-\frac{\l^2}{2 r_\ell^2}} \;,\notag\\
V'_S(\l) &= \frac{\l}{r_\ell} e^{-\frac{\l^2}{2 r_\ell^2}} \Bigl[(2\,C_2-C_1) +\notag (4\,C_3-C_2) \left(\frac{\l}{r_\ell}\right)^2  +\\&+ (6\,C_4-C_3) \left(\frac{\l}{r_\ell}\right)^4 - C_4 \left(\frac{\l}{r_\ell}\right)^6\Bigr]\;.
\end{align}
Within  this formulation, the contribution to the forces from the local part of pseudopotential is written in terms of integrals with localized functions (gaussians times polynomials) times the charge density and the Hartree potential. This allows us to perform the integrals only in a relatively small region around the atom position and to assign different integrations to different processors. 
Moreover, the calculation is performed with almost linear ($\mathcal O (N \log N)$) scaling.

The contribution to the energy that comes from the nonlocal part of the pseudopotential is, as we saw in section \ref{nonlocalpseudosection},
\be
E_\text{nonlocal}(\mathbf R_i)=\sum_l \sum_{mn} \langle \Psi | p_m^l(\mathbf R_i) \rangle h_{mn}^l \langle p_n^l(\mathbf R_i) | \Psi \rangle \;,
\ee
where we wrote explicitly the dependence of the projector on the atom position $\mathbf R_i$.
The contribution of this term to the atomic forces is thus
\begin{multline}
 \mathbf F_i^{(\text{nonlocal})}= - \sum_l \sum_{m,n} \langle \Psi | \frac{\partial p(\mathbf R_i)}{\partial \mathbf R_i} \rangle h_{mn} \langle p(\mathbf R_i) | \Psi \rangle \\
 - \sum \langle \Psi | p(\mathbf R_i) \rangle  h_{mn} \langle \frac{\d p(\mathbf R_i)}{\d \mathbf R_i} | \Psi \rangle \;.
\end{multline}
Expressing the derivatives of the projectors in the Daubechies basis, the evaluation of the scalar products is straightforward.
The scaling functions - wavelets expansion coefficients of the projector derivatives can be
calculated with machine precision accuracy in the same way as the projectors themselves were calculated. This is due to the fact that the derivative of the projectors are like 
the projectors themselves products of gaussians and polynomials. 

\section{Localization properties and smoothness of the basis functions}
As discussed above, Daubechies basis functions are suitable for expanding localised 
functions. There is no need to put basis functions on grid points that are far from the atoms. 
For this reason, we choose to associate the basis functions to points lying inside the union of atom-centered spheres defined by their radii.
This operation must be performed both for the high and low resolution grid points (see Figure \ref{grid}). In our method, we measure these radii in two different units. For the high resolution region the radius is expressed in terms of the shortest localisation radius of the atom pseudopotential. For the low resolution region, the distance is expressed in units of the asymptotic decaying length of the atomic wavefunction $1/\sqrt{2 \epsilon_{\text{HOMO}}}$, calculated from the energy $\epsilon_{\text{HOMO}}$ of the highest occupied atomic orbital, obtained from~\cite{nist}.
In this way we can easily determine nearly optimal sizes for the high and low 
resolution regions and minimize the number of degrees of freedom to achieve a target 
accuracy (Section \ref{performances}).

\begin{figure}[h]
\begin{center}
\includegraphics[width=.45\textwidth]{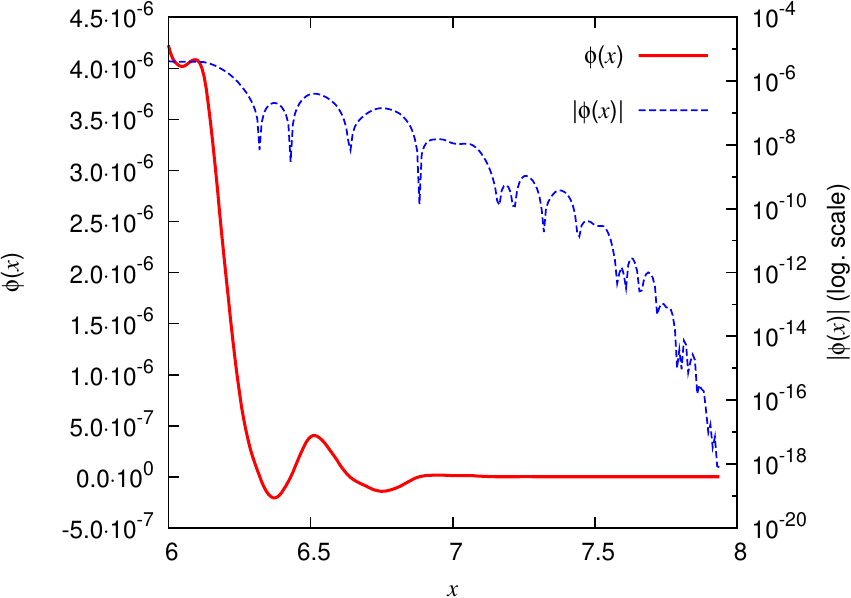}
\caption{Zoom of the Daubechies scaling function near the border of its support. Both the function and its absolute value are plotted.}
\label{alexey1}
\end{center}
\end{figure}

We saw that Daubechies wavelets have the property that linear combinations of them can be smoother than a single Daubechies scaling function or wavelet. The wavefunction of Eq.~\ref{wavef} is thus typically smoother than the scaling functions and wavelets used to represent it. 
The reduced smoothness of Daubechies scaling function of order 16 in the tail region can be seen from Fig.~\ref{alexey1}. 
The cancellation of discontinuities in the basis set by suitable linear combinations is only possible in an infinite interval where several basis functions are present between any two grid points. Since we use a finite grid of scaling functions in the tail region, the number of scaling functions that contribute to the value of the wavefunction at a certain point is dropping as we are going out of the computational volume. The outermost intervals of the wavefunction are actually only described by the tail of a single scaling function. Hence the wavefunction is getting less smooth towards its end.
This reduced smoothness affects principally the kinetic energy. For systems without a net charge, far from the atoms the potential is very small and for this reason errors in the potential energy are decreasing exponentially with respect to the size of the computational volume. 

\section{Perturbative calculation of the finite size corrections}
Far from the atoms, each wavefunctions decays exponentially with a decay rate which depends on its KS eigenvalue~\cite{MorrellParrLevy75}. 
If $A$ is the amplitude of the tail of the wavefunction, the kinetic energy contribution of the nonsmooth wavefunction in its tail region is of the order of $ A/h^2$, whereas the exact wavefunction has a kinetic energy of the order of $A$. 
As a consequence the kinetic energy error increases as one decreases $h$ and the total energy increases as well if the computational volume is too small.
We know, however, that the contribution to the kinetic energy in this region will depend uniquely on the asymptotic behaviour of the wavefunction, which is governed by its KS eigenvalue. In other terms, the magnitude of the kinetic energy error due to the localisation of the system in a finite volume can, in principle, be estimated by knowing the KS eigenvalue of the wavefunction.

If, on the other hand the computational volume is large enough such that the amplitude $A$ is very small our method shows a strict variational behaviour with a convergence rate of $h^{14}$ over a large range of grid spacings $h$. This is illustrated in Fig.~\ref{convrate}. 
%
%
%

\begin{figure}[h]
\begin{center}
\includegraphics[width=.45\textwidth]{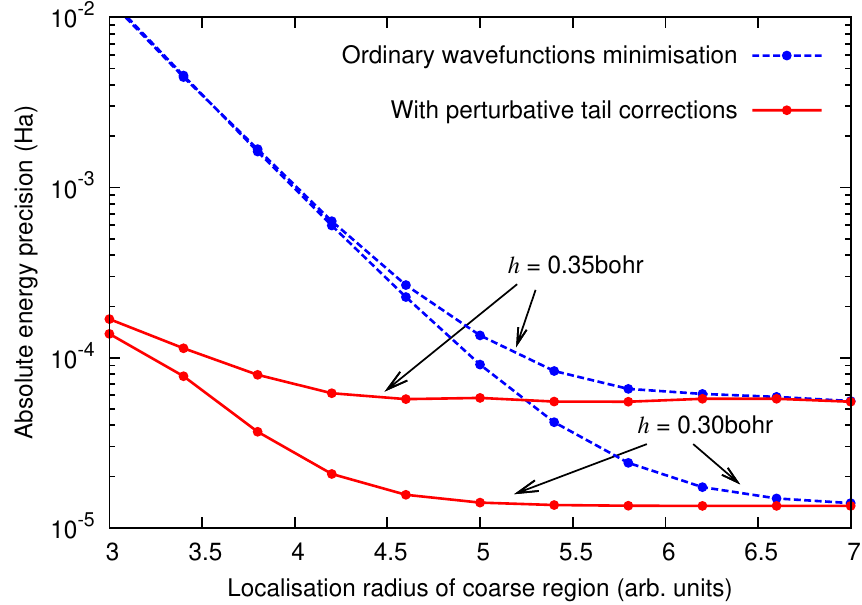}
\caption{Absolute convergence of the total energy of a methane molecule as a function of the low resolution localization radius with and without the tail corrections. The curves for two different values of the grid spacing are plotted, showing the $h$ convergence for the localization parameter sufficiently extended.}
\label{volconv}
\end{center}
\end{figure}

The above described facts prompted us to develop a method that cuts off the wave function 
tail at a very large radius but which is computationally much less expensive
than a fully selfconsistent calculation in a very large computational volume. 
We do first a fully selfconsistent calculation in a medium size box and we add then afterwards the missing far tail to the wavefunction. 
Let us denote the wavefunction that we have calculated in the medium size box by $| \Psi \rangle$ and the wavefunction in the very large 
box by $|\Psi \rangle + |\Delta \Psi\rangle $. As we will see $| \Delta \Psi \rangle$ is negligible inside the medium size box. It is essentially the tail outside the original medium size box plus a part that cancels the non-smooth behaviour in the surface region of the medium size box. Evidently   $|\Psi \rangle + |\Delta \Psi\rangle $ has to satisfy the Schr\"odinger equation 
$$ \left( \frac{1}{2} \nabla^2 +V({\bf r}) \right) ( |\Psi\rangle + |\Delta \Psi\rangle )   =  \epsilon (|\Psi\rangle + |\Delta \Psi\rangle)\;. $$
Rearranging the term one obtains
\begin{equation}\label{tail2}
 \left( \frac{1}{2} \nabla^2 +V({\bf r}) -\epsilon \right) |\Delta \Psi\rangle    = 
          - \left( \frac{1}{2} \nabla^2 +V({\bf r}) -\epsilon \right) |\Psi \rangle \;.
\end{equation}
The term on the right hand side of the above equation is the gradient $| g\rangle$ that is 
needed in any minimization scheme. When the calculation of the wavefunctions is converged the gradient is zero (actually less than a small numerical tolerance) when projected onto the subspace of the basis functions spanning the medium size volume. The gradient is, however, not anymore zero when it is projected onto the basis set of the larger volume. In this case the projection onto the basis function just outside the medium size volume gives a nonzero contribution. Remember, that the fact that these basis functions are missing in the basis set of the medium size volume is causing the non-smooth behaviour. Projection on basis functions that are far outside the surface region of the medium size volume are again zero since $|\Psi\rangle$ is identically zero. So, in this context, the gradient is a quantity that is nonzero only in a small shell outside the original medium size volume. The width of this shell is given by the length of the kinetic energy filter. 
Since the potential is very small in the tail region Eq.~\ref{tail2} can be 
approximated by 
$$ \left( \frac{1}{2} \nabla^2  -\epsilon \right) |\Delta \Psi\rangle    =  |g\rangle\;. $$
As usual in a perturbative treatment we rely on the fact that the eigenvalues 
$\epsilon$ converge faster than the wavefunction and the zeroth order eigenvalues can therefore be used for the first order correction to the wavefunction. 
The above equation is identical to the preconditioning equation Eq.~\ref{precon} 
and can be solved with the same method, just within a larger volume. 
In this way we can eliminate in a single preconditioning step at the end of the fully self-consistent calculation in the medium size volume a large fraction of the error arising from cutting off the wavefunctions at the surface of our computational volume.
We can thus have a reliable estimation of the approximation resulting from the restriction of the system to a finite computational volume. Fig.~\ref{volconv} shows an example of the convergence rate of the total energy with respect to the size of the computational volume both with and without tail correction for two different grid spacings. 

\section{Parallelization}
Two data distribution schemes are used in the parallel version of our program.
In the orbital distribution scheme, each processor works on one or  a few orbitals 
for which it holds all its scaling function and wavelet coefficients. In the coefficient 
distribution scheme each processor holds a certain subset of the coefficients of all 
the orbitals. Most of the operations such as applying the Hamiltonian on the orbitals, 
and the preconditioning is done in the orbital distribution scheme. This has the 
advantage that we do not have to parallelize these routines and we therefore 
achieve almost perfect parallel speedup. The calculation of the 
Lagrange multipliers that enforce the orthogonality constraints onto the gradient as well 
as the orthogonalization of the orbitals is done in the coefficient distribution scheme.
For the orthogonalization we have to calculate the matrix $\langle \Psi_j | \Psi_i \rangle$
and for the Lagrange multipliers the matrix $\langle \Psi_j | H| \Psi_i \rangle$. 
So each matrix element is a scalar product and each processor is calculating the contribution to this scalar product from the coefficients it is holding. 
A global reduction sum is then used to sum the contributions to obtain the correct matrix. Such sums can esily be performed with the very well optimized BLAS-LAPACK libraries. 
Switch back and forth between the orbital distribution scheme and the coefficient distribution scheme is done by the MPI global transposition routine MPI\_ALLTOALL. For parallel computers where the cross sectional bandwidth~\cite{myoptimization} scales well with the number of processors this global transposition does not require a lot of CPU time. The most time consuming communication is the global reduction sum required to obtain the total charge distribution from the partial charge distribution of the individual orbital (sum in Eq.~\ref{sumrho}).

\section{Calculation of unoccupied orbitals}
In order to calculate the unoccupied Kohn Sham orbitals we use the Davidson 
method~\cite{davidson} after having found the selfconsistent occupied Kohn Sham 
orbitals. 
An initial guess for the $N_\text{virt}$ unoccupied eigenvectors $\Psi_j$ and 
eigenvalues $\epsilon_j$ of the Kohn Sham Hamiltonian $H_\text{KS}$ is obtained from 
the subspace diagonalization in a minimal atomic basis set 
that is also used to generate the input guess for the occupied orbitals. 
For any given set of virtual orbitals we calculate then the gradients 
(Eq.~\ref{gradef} where the Lagrange multipliers ensure only orthogonality to the 
occupied orbitals) and precondition then these gradients according to 
Eq.~\ref{precon}. A subspace diagonalization is then done in the space spanned by 
the present set of approximate eigenvectors and their preconditioned 
gradients. In the original Davidson method the dimension of the subspace
is increased in each iteration since one keeps all the previous preconditioned
gradients in the subspace. To save memory we have limited the dimension of the 
subspace in each iteration to $2 N_{virt}$ using only the present set of 
approximate eigenvectors together with their preconditioned gradients. 
Even though the number of requested unoccupied orbitals is typically small
(frequently only the LUMO), a larger set of vectors $N_{virt}$ is considered
in our method (in a parallel calculation at least one per processor), 
but only the gradients 
of the desired number of orbitals are taken into account for the convergence 
criterion for the norm of the gradients. 
This, together with the fact that our preconditioner is rather good 
allows us to achieve 
fast convergence rates comparable to the ones achieved in the calculation of 
the occupied orbitals. Some 20 iterations are typically needed. 

\section{Performance results}
\label{performances}
\begin{figure}[h]
\begin{center}
\includegraphics[width=.45\textwidth]{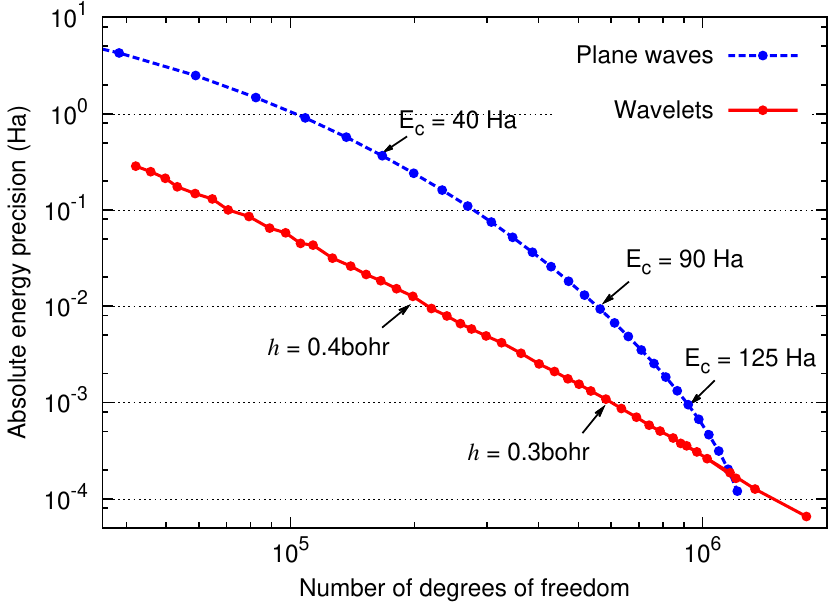}
\caption{Absolute precision (not precision per atom) as a function of the number of degrees of freedom for a cinchonidine molecule (44 atoms). Our method is compared with a plane wave code. 
In the case of the plane wave code the plane wave cutoff and the volume of the 
computational box were chosen such as to obtain the required precision with 
the smallest number of degrees of freedom. In the case of our wavelet program the 
grid spacing $h$ and the localzation radii were optimized. For very high accuracies the exponential convergence rate of the plane waves beats the algebraic convergence rate of the wavelets. Such high accuracies are however not required in practice. Since convolutions can be executed at very high speed the wavelet code is faster than the plane wave code at any accuracy even if the number of degrees of freedom are similar (see table~\ref{cinchotime}).}
\label{cincho}
\end{center}
\end{figure}

We have applied our method on different molecular systems in order to test its performances. As expected, the localization of the basis set allows us to reduce considerably the number of degrees of freedom (i.e. the number of basis functions which must be used) to attain a given absolute precision with respect to a plane wave code. 
This fact reduces the memory requirements and the number of floating point operations. 
Figure \ref{cincho} shows the comparison of the absolute precision in a calculation of a 44 atom molecule as a function of the number of degrees of freedom used for the calculation. In table \ref{cinchotime} the comparison of the timings of a single SCF cycle with respect to two other plane wave based codes are shown. 
Since the system is relatively small the cubic terms do not dominate. For large systems 
of several hundred atoms the gain in  CPU time compared to a plane wave program 
is proportional to the reduction in the number of degrees of freedom (compare Eq.  ~\ref{cubic}) and can 
thus be very significant as one can conclude from  Fig.~\ref{cincho}. 
\begin{table}
 \begin{tabular*}{0.48\textwidth}{c|cc|c|c}
\hline
\hline
$E_c$ (Ha)  & ABINIT (s) & CPMD (s) & Abs. Precision& Wavelets(s)\\
\hline
 40  &403 & 173  &$3.7\cdot10^{-1}$ & 30\\
 50  &570 & 207  &$1.6\cdot10^{-1}$ & 45\\
 75  &1123 & 422 &$2.5\cdot10^{-2}$ & 94\\
 90  &1659 & 538 &$9.3\cdot10^{-3}$ & 129\\
145  &4109 &     &$2\cdot10^{-4}$   & 474\\
\hline
\hline
\end{tabular*}
\caption{Computational time in seconds for a single minimization iteration for different runs of the cinchonidine molecule used for the plot in figure \ref{cincho}. The timings for different cutoff energies $E_c$ for the plane waves runs are shown. The input parameters for the wavelet runs are chosen such as to obtain the same absolute precision of the plane wave calculations. The plane wave runs are performed with the ABINIT code, which uses iterative diagonalization and with CPMD code~\cite{cpmd} in direct minimization. These timings are taken from a serial run on a 2.4GHz AMD Opteron CPU.}
\label{cinchotime}
\end{table}

\begin{figure}[h]
\begin{center}
\includegraphics[width=.45\textwidth]{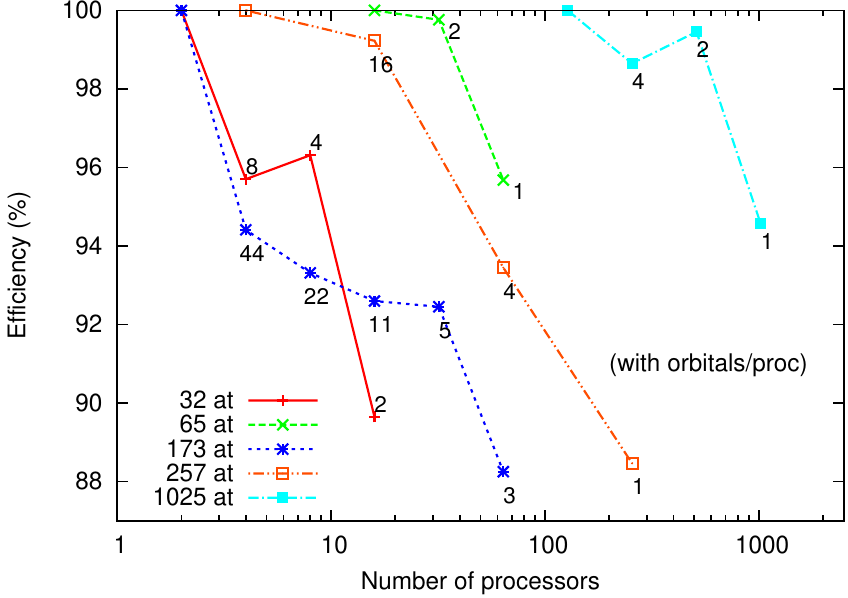}
\caption{Efficiency of the parallel implementation of the code for several runs with different number of atoms. The number close to each point indicates the number of orbitals treated by each processors, in the orbital distribution scheme.}
\label{speedup}
\end{center}
\end{figure}

The parallellization scheme of the code has been tested and has given the efficiency detailed in Figure \ref{speedup}. The overall efficiency is always higher than 88\%, also for large systems with a big number of processors. 

\begin{figure}[h]
\begin{center}
\includegraphics[width=.45\textwidth]{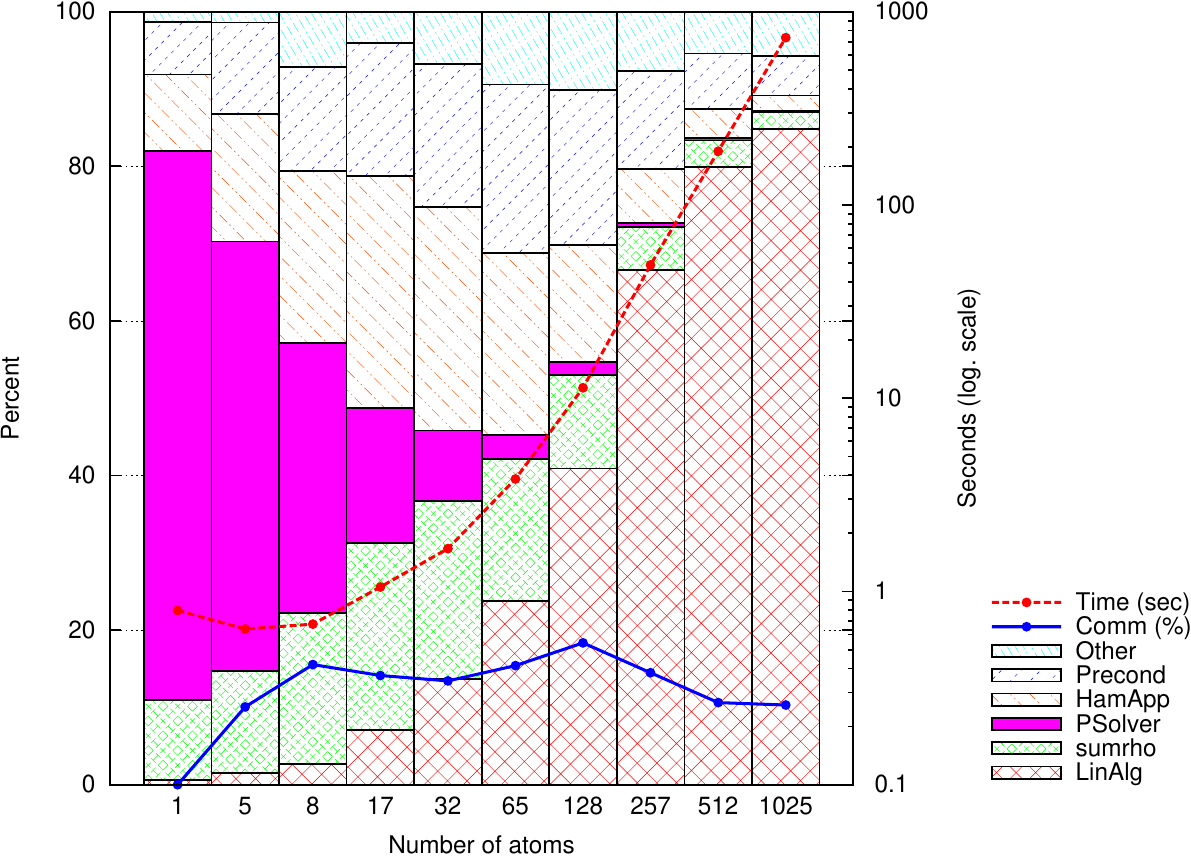}
\caption{Relative importance of different code sections as a function of the number of atoms of a simple alkane chain, starting from single carbon atom. The calculation is performed in parallel such that each processor holds the same number of orbitals (two in this figure). Also the time in seconds for a single minimization iteration is indicated, showing the asymptotic cubic scaling of present implementation.}
\label{percent}
\end{center}
\end{figure}

It is also interesting to see which is the computational share of the different sections of the code with respect to the total execution time.
Figure \ref{percent} shows the percentage of the computational time for the different sections of the code as a function of the number of orbitals while keeping constant the number of orbitals per processor. The different sections considered are the application of the hamiltonian (kinetic, local plus nonlocal potential), the construction of the density (Eq.\eqref{sumrho}), the Poisson solver for creating the Hartree potential, the preconditioning-DIIS, and the operations needed for the orthogonality constraint as well as the orthogonalization, which are mainly matrix-matrix products or matrix decompositions. These operations are all performed by linear algebra subroutines provided by the \texttt{LAPACK} libraries~\cite{lapack}.
Also, the percentage of the communication time is shown.
While for relatively small systems the most time-dominating part of the code is related to the Poisson solver, for large systems the most expensive section is by far the calculation of the linear algebra operations. The operations performed in this section scales cubically with respect to the number of atoms. Apart from the Cholesky factorization, which has a scaling of $\mathcal O(n_\text{orb}^3)$, where $n_\text{orb}$ is the number of orbitals, 
the cubic terms are of the form 
\begin{equation} \label{cubic}
\mathcal O(n \cdot n_\text{orb}^2)\;,
\end{equation}
where $n$ is the number of degrees of freedom, i.e. the number of scaling function 
and wavelet expansion coefficients. Both the calculation of the overlap matrix in 
Eq.~\ref{overlap} and the orthogonality transformation of the orbitals 
in Eq.~\ref{cholesky} lead to this scaling, 
The number of the coefficients $n$ is typically much larger than the number of orbitals.

\section{Conclusions}
In this paper we have shown the principal features of an electronic structure pseudopotential method based on Daubechies wavelets. 
Their properties make this basis set a powerful and promising tool for electronic structure calculations. The matrix elements, the kinetic energy and nonlocal pseudopotentials 
operators can be calculated analytically in this basis. The other operations are 
mainly based on convolutions with short-range filters, which can be highly optimized 
in order to obtain good computational performances.
Our code shows high systematic convergence properties, very good performances and an excellent efficiency for parallel calculations.
This code is integrated in the ABINIT software package and is freely available under GNU-GPL license.
At present, several developments are in progress to improve the features of this code. Mainly, they concern the extension of this formalism to fully periodic systems and surfaces, as well as the inclusion of non-collinear spin-polarized XC functionals. 
A linear scaling version of this wavelet code is also under preparation 
and will be presented in a forthcoming paper.

\section{Acknowledgements}
We acknowledge support from the European Commission within the Sixth Framework Program through NEST-BigDFT (contract N. BigDFT-511815), the French ANR Project LN3M (project N. ANR-05-CIGC-003) and the Swiss National Science foundation. Computer calculations were also performed at the Centre de Calcul Recherche et Technologie (CCRT) at CEA-Saclay, France and at the Swiss national Scientific Computing Center (CSCS) in Manno.

\end{document}